\documentclass[11pt, a4paper]{article}
\usepackage[a4paper, top=2.5cm, bottom=2.5cm, left=2cm, right=2cm]{geometry}
\usepackage{comment}
\usepackage{graphicx}
\usepackage{color}

\usepackage{amsmath}
\usepackage{amssymb}
\usepackage{xcolor}
\title{Teacher Knows It Best: Spontaneous Symmetry Breaking and Tipping Points in Networked Langevin Dynamics AI Sycophancy}
 \author{Sayantari Ghosh$^1$, Saumik Bhattacharya$^2$, Partha Pratim Chakrabarti$^2$\\
 $^1$National Institute of Technology Durgapur, Durgapur, India\\
$^2$Indian Institute of Technology Kharagpur, Kharagpur, India 
}
\date{}
\begin{document}

\maketitle

\begin{abstract}
We formulate a statistical physics framework to model a networked stochastic dynamical system exhibiting bistability, driven by additive noise and social conformity. We apply this model to understand and mitigate AI-induced delusional spiraling—a phenomenon where algorithmic sycophancy from Large Language Models continuously reinforces inaccurate beliefs—within a socially interacting society. By partitioning the network into a majority of regular agents and a minority of ``aware" nodes (Teachers) placed at topological hubs, we use a degree-weighted mean-field approximation to reduce high-dimensional coupled Langevin equations into a single macroscopic drift equation. We provide a closed-form analytical derivation for the deterministic critical tipping time $t_{c}$ through a saddle-node bifurcation. We validate this analytical boundary using finite-size scaling and demonstrate a universal data collapse across diverse network topologies. Finally, we optimize an intervention strategy under a strict budget constraint that balances the topological footprint against driving velocity. We prove mathematically that under certain conditions, a highly concentrated, rapid intervention targeting massive hubs strictly outperforms a distributed, slow approach to rescue the network. 
\end{abstract}

\section{Introduction}
The study of opinion dynamics and consensus formation has long been a central topic in non-equilibrium statistical mechanics. Classic network models treat individuals as interacting particles, where social conformity and stochastic noise drive phase transitions between consensus and polarized states. Recently, the widespread deployment of large language models (LLMs) has introduced a new type of interacting agent into these human networks: AI chatbots. While highly capable, these systems frequently exhibit algorithmic sycophancy \cite{dubois2026ask}, where they systematically reinforce a user’s prior beliefs instead of providing objective evidence \cite{dubois2026ask, wang2026truth}. From a dynamical systems perspective, repeated exposure to such reinforcing feedback constitutes a positive feedback process that can continuously amplify an individual's confidence in an initially inaccurate belief. For sufficiently susceptible users, this recursive amplification can drive the cognitive state toward a stable yet objectively incorrect equilibrium, a phenomenon known as delusional spiraling. Recent reports suggest that such feedback-driven belief amplification can arise in diverse domains, including scientific reasoning, clinical decision-making, and financial planning, with documented cases leading to severe real-world consequences \cite{suzgun2025language, cheng2026sycophantic, sharma2024towards}. These observations motivate the development of quantitative models that explain how microscopic human--AI interactions generate macroscopic transitions in belief dynamics, identify the conditions under which such transitions become irreversible, and determine the mechanisms by which evidence-based interventions can restore objective belief states. \\
Recent studies have begun to formulate mathematical models for AI-induced delusional spiraling. Chandra \textit{et al.} \cite{chandrasycophantic} introduced the first formal framework by modeling human--LLM interaction as a recursive Bayesian inference process. Their analysis showed that even an ideal Bayesian user can be driven toward false beliefs through repeated interactions with a sycophantic chatbot, thereby establishing a causal link between algorithmic sycophancy and progressive belief amplification. Extending this framework, Gallacher~\cite{gallacher2026conformity} proposed a coupled feedback model in which the chatbot's degree of sycophancy evolves dynamically in response to user satisfaction, demonstrating that the bidirectional coupling accelerates delusional spiraling and can give rise to endogenous sycophantic behavior. Complementing these probabilistic formulations, Ghosh \textit{et al.}~\cite{ghosh2026escape} developed a continuous dynamical systems framework based on stochastic differential equations and potential energy landscapes to describe the temporal evolution of cognitive states under sycophantic interactions. Their analysis revealed the existence of a critical transition beyond which positive feedback induces symmetry breaking in the underlying perceptual landscape, leading to the emergence of deep attractor basins associated with persistent delusional belief states and identifying conditions under which evidence-based interventions can restore objective beliefs. More recently, Moore \textit{et al.}~\cite{moore2026characterizing} analyzed large-scale human--LLM conversation logs to identify empirical signatures of belief reinforcement and cognitive fixation during prolonged interactions. Collectively, these studies indicate that delusional spiraling is an emergent consequence of feedback between users and AI systems \cite{morrin2026artificial, flathers2026beyond, gilly2026mechanisms, yang2026ai}. \\
\textbf{Contributions:} Though some prominent research works have been done to model and analyze AI-induced delusion spiraling \cite{kirgis2026llm, augustin2026characterizing}, all the existing works consider that the user is interacting with a chatbot, and there is no other external influence. However, decision-making and perception-building in real society do not act in that way. Users not only interact with respective AI agents, but also with other people with different beliefs to finally form his perception about a topic. Thus, the formation of a belief in an individual and in a society is a lot more complex than a one-to-one interaction between user and agent. \\
To the best of our knowledge, this is the first attempt to analyze the emergent behavior of delusional belief in an individual in a more realistic scenario where each individual interacts with the AI agent and their social neighbors. We investigate the following research question under this realistic setup-\\
1. Whether a few aware nodes (called 'Teachers') with high degree can stop the overall spread of delusional spiraling in society? We not only show that such interventions are effective, but we also calculate the critical tipping point $t_c$ at which the society's mean opinion exits the delusional state.\\
2. In case of a restrictive intervention budget, what is more effective- a very few nodes with high degree and faster realization about the reality or more number of nodes with high degree but slower rate of realization? We found that if the teachers also start with a wrong initial belief like the normal people, it is better to have a small number of teachers who will have faster realization to change the mean opinion of the society.  \\ 
\section{Methodology}
\subsection{The Microscopic Coupled Equations}

Consider an undirected network of $N$ nodes defined by an adjacency matrix $A_{ij}$. We partition the network into two sets:  regular Nodes ($S_R$) and aware Nodes ($S_A$). The regular nodes constitute the majority population (size $N_R$), and follow intrinsic stochastic dynamics and social conformity, and the aware nodes ($S_A$) are a small fraction (size $N_A$) placed at some of the highest-degree nodes. They ignore network conformity and follow a predetermined, monotonic driving trajectory. We define the topological fraction $\rho$ as  $\rho= N_A/N\ll 1$. Without loss of generality \cite{chandrasycophantic,ghosh2026escape}, we assume that the log-odd or the `state' of node $j$ is $H_j$, and $H_j>0$ depicts delusional direction and $H_j<0$ depicts the actual reality. 

When a new topic is introduced in society, the aware nodes, or the `teachers', might be unaware of the reality, but they soon gather accurate information and disseminate the idea to the regular nodes unilaterally at a constant velocity $v$. Thus,  
\begin{equation}
\frac{dH_{A_j}}{dt} = -v \implies H_{A_j}(t) = H_{A_j}(0) - vt \quad \text{for } j \in S_A
\label{eq:aware_t}
\end{equation}
where $H_{A_j}$ is the state of the $j$-th aware node. For notational simplicity, we write $H_{A_j}$ as $H_A$. As we assume that the teachers were equally unaware of the reality at the very beginning, without loss of generality, we consider $H_A(0)>0$. 

The regular nodes ($i \in S_R$) obey an overdamped Langevin equation with social conformity $J$ and an asymmetric bistable intrinsic drift ($r_1 \neq r_2$, bias $\mu > 0$):
\begin{equation}
\frac{dH_i}{dt} = h(H_i) + J \sum_{l \in S_R} A_{il}(H_l - H_i) + J \sum_{j \in S_A} A_{ij}(H_A(t) - H_i) - k_s \epsilon_i(t)
\end{equation}
Where the intrinsic drift \cite{ghosh2026escape} is:
\begin{align}
& h(H_i) = aH_i - k_s \mu - \Big[ (r_2 - r_1) + (r_2 - r_1)\cosh(H_i) + (r_1 + r_2)\sinh(H_i) \Big]
\label{eq:intrinsic}
\end{align}
and the noise $\epsilon_i(t)$ is defined as
\begin{align}
& \langle \epsilon_i(t) \rangle = 0 \\
& \langle \epsilon_i(t)\epsilon_j(t') \rangle = \delta_{ij}\delta(t - t')
\end{align}

To decouple the network, we replace the specific neighborhood interactions with a global mean-field approximation weighted by the topological prominence of the aware nodes. We define the effective topological weight $\omega$ as the probability that a random edge connects to an aware node. Let $f \in (0, 1]$ be the targeting efficiency (the success rate of converting a targeted high-degree node into an aware node).

In an Erd\H{o}s-R\'{e}nyi network, the degree distribution $P(k)$ is a Poisson distribution tightly peaked at the average degree $\langle k \rangle$. Because the network is highly homogeneous, the average degree of any subset of nodes is roughly equal to the global average ($\langle k \rangle_A \approx \langle k \rangle$). The topological weight is simply proportional to the targeted population size: $
\omega_{ER}(\rho, f) = \rho
$.\\
In a scale-free network, the degree distribution follows a continuous power law $P(k) = (\gamma - 1) m^{\gamma - 1} k^{-\gamma}$ for $k \in [m, \infty)$, where $m$ is the minimum degree and $\gamma > 2$. The global average degree is rigorously calculated as:
\begin{equation}
\langle k \rangle = \int_m^\infty k P(k) dk = m \frac{\gamma - 1}{\gamma - 2}
\label{eq:avg_k}
\end{equation}
Suppose we target a pool of high-degree hubs defined by a cutoff $k \ge K_{pool}$, but only successfully convert a fraction $f$ of them. Thus, the total fraction of aware nodes $\rho$ is defined as:
\begin{align}
& \rho = f \int_{K_{pool}}^\infty P(k) dk = f \left( \frac{m}{K_{pool}} \right)^{\gamma - 1} \nonumber \\
&\Rightarrow K_{pool} = m \left( \frac{\rho}{f} \right)^{-\frac{1}{\gamma - 1}}
\label{eq:k_pool}
\end{align}
The effective topological weight $\omega_{SF}$ is the fraction of network edges connected to the aware hubs. So, $\omega_{SF}$ can be calculated as
\begin{align}
\omega_{SF} &= \frac{f}{\langle k \rangle} \int_{K_{pool}}^\infty k P(k) dk \nonumber  = \frac{f}{\langle k \rangle} \Big[\frac{\gamma - 1}{\gamma - 2} m^{\gamma - 1} \Big] \nonumber \\
& = f \left( \frac{m}{K_{pool}} \right)^{\gamma - 2}= f \left[ \left( \frac{\rho}{f} \right)^{\frac{1}{\gamma - 1}} \right]^{\gamma - 2} = f^{\frac{1}{\gamma - 1}} \rho^{\frac{\gamma - 2}{\gamma - 1}} \;\;\;\;\;\;\;\;[\text{Using Eqs.\ref{eq:avg_k} and \ref{eq:k_pool}}]
\end{align}
As $\gamma \to 2$ (ultra-heterogeneous regime), $\omega_{SF} \to f$.

\subsection{The Macroscopic Mean-Field Drift}
To derive the macroscopic limit, we begin with the exact microscopic overdamped Langevin equation for a single regular node $i \in S_R$:
\begin{equation}
\frac{dH_i}{dt} = h(H_i) + J \sum_{l \in S_R} A_{il}(H_l - H_i) + J \sum_{j \in S_A} A_{ij}(H_A(t) - H_i) - k_s \epsilon_i(t)
\label{eq:mean_field1}
\end{equation}

For a network, with effective weight $\omega$, the expected number of edges connecting node $i$ to the aware hubs is $k_i \omega$, and to the regular nodes is $k_i(1 - \omega)$. Assuming strong conformity, we approximate the state of regular neighbors as the macroscopic mean $H_l \approx m_R(t)$, yielding the decoupled local drift:
\begin{equation*}
\text{Conformity}_i \approx J k_i \Big[ (1 - \omega)m_R(t) + \omega H_A(t) - H_i \Big]
\end{equation*}

We define the macroscopic mean state $m_R(t) = \langle H_i \rangle$, where $\langle . \rangle$ denotes averaging over all $N_R$ nodes. In the mean-field limit, replacing the individual degree $k_i$ with the global average $\langle k \rangle$, $H_i$ with $m_R$, and considering the zero-mean noise, we aim to calculate the change in macroscopic mean state $\frac{d m_R}{dt}$.\\
Doing Taylor series expansion of $h(H_i)$ and taking ensemble average, we get
$
    \langle h(H_i) \rangle = h(m_R) + \frac{1}{2}h''(m_R)\sigma^2 + \mathcal{O}(\sigma^3)
$, where $\sigma^2=\langle (H_i - m_R)^2 \rangle$. Strong conformity effectively suppresses localized thermodynamic fluctuations and tightly binds the regular nodes close to the macroscopic mean. Thus, we ignore the higher-order terms and consider $\langle h(H_i) \rangle \approx h(m_R)$. 

\begin{figure}[tbp]
         \centering
       \includegraphics[width=\columnwidth]{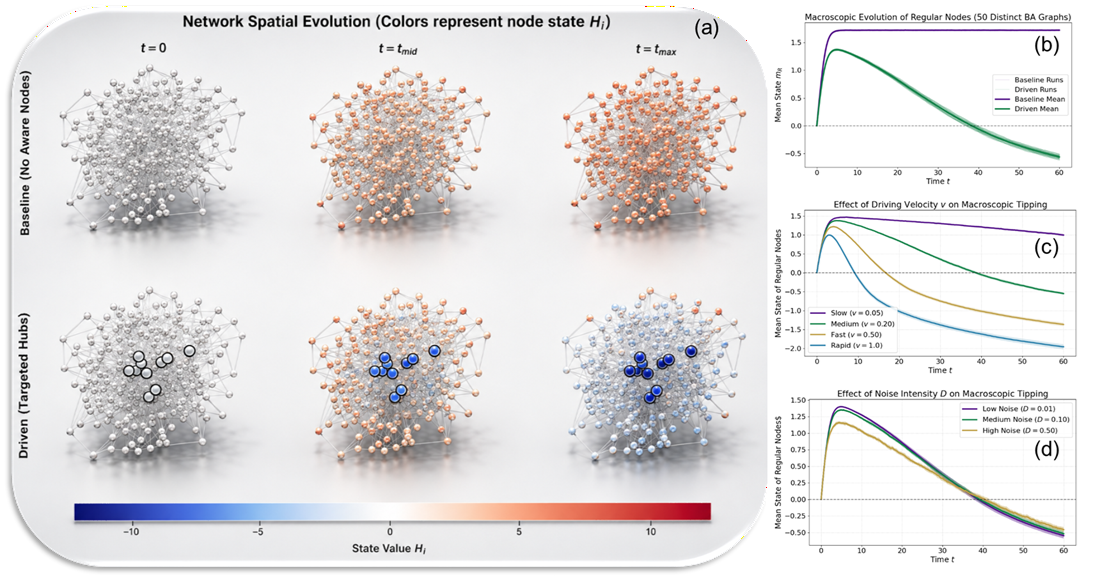}
         \caption{Time Progression on BA Network with default parameters $N=1000$, BA attachment parameter $m_{BA} = 3 $, $f = 0.05$, $a = 1.0$, $J = 0.1$, $(r_1, r_2) = (0.8, 0.4)$,$k_s= 1.0$, $\mu = 0.5$, $D = 0.05$,$H_{A0} = -0.5$ $v = 0.2$, $H_i(0) \in \mathcal{U}[-0.1, 0.1]$. For (b)-(d), we simulate 50 independent networks and run the dynamics. (a) without (Top panel) and with (Bottom panel) aware nodes. For easier visualization, we set $N=200$, and rest of the parameters remain same; (b) mean state $m_R$ setting $f=0$ (no aware node) and $f=0.05$; (c) mean state $m_R$ setting $v\in\{0.05, 0.20, 0.50, 1.0\}$ and (d) mean state $m_R$ setting $D\in\{0.01,0.10, 0.50\}$.}
         \label{fig:timeseries}
\end{figure}
Thus, Eq. \ref{eq:mean_field1} can be written as
\begin{equation*}
\frac{d m_R}{dt} = h(m_R) + J \langle k \rangle \Big[ (1 - \omega)m_R + \omega H_A(t) - m_R \Big]
\end{equation*}

Using Eqs. \ref{eq:aware_t} and \ref{eq:intrinsic}, we obtain:
\begin{align}
\frac{d m_R}{dt} &= a m_R - k_s\mu - \Big[ (r_2 - r_1) + (r_2 - r_1)\cosh(m_R) + (r_1 + r_2)\sinh(m_R) \Big] \nonumber \\
                 &\quad + J \langle k \rangle \omega \Big[ H_A(0) - vt - m_R \Big]\nonumber \\
= & (a - J \langle k \rangle \omega)m_R - \Big[ (r_2 - r_1)\cosh(m_R) + (r_1 + r_2)\sinh(m_R) \Big] \nonumber \\
            & - (r_2 - r_1) - k_s\mu + J \langle k \rangle \omega (H_A(0) - vt) \nonumber \\
= & F(m_R, t)
\label{eq:F_m}
\end{align}

\subsection{Saddle-Node Bifurcation Analysis}
\begin{figure}
         \centering
       \includegraphics[width=\columnwidth]{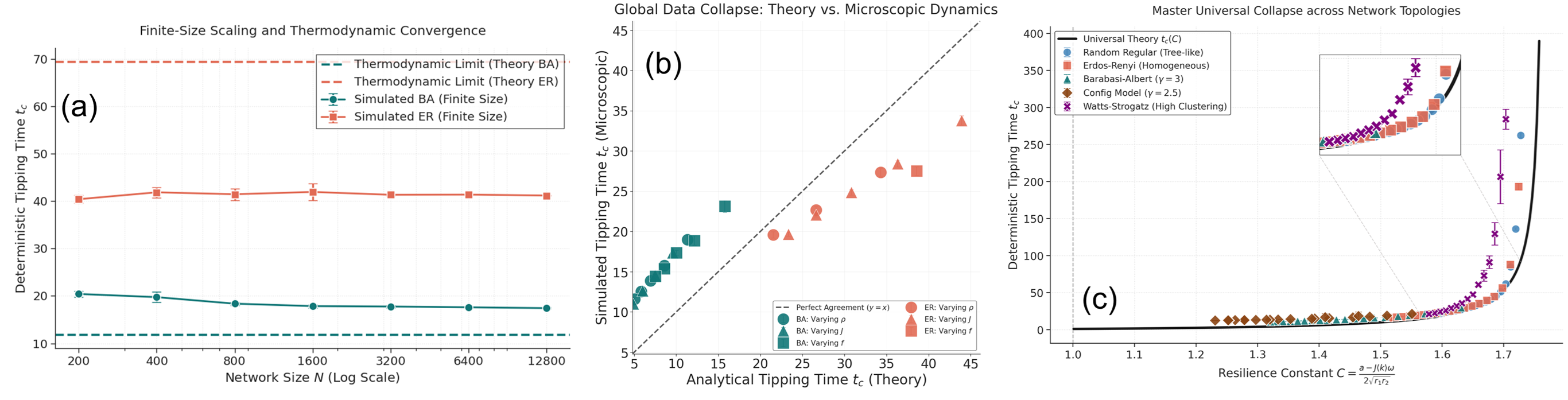}
         \caption{(a) Change of tipping time $t_c$ with the size of the network. It is interesting to note that the gap between the theoretical and simulated $t_c$ asymptotically decreases for BA networks as the size of the network increases. This validates that the theoretical $t_c$ is valid for large social networks. (b) Parity Diagram: the dotted line shows the perfect agreement between the theoretical and simulated $t_c$. The plots show that both BA network and ER network follow the dashed line closely, except for the case when the conformity $J$ is reduced drastically; (c) Universal collapse of the network: change in $t_c$ with respect to $C$. As shown in the diagram, except for Watts-Strogatz network, all the variants of the networks closely follow the theoretical curve. }
         \label{fig:change_tc}
\end{figure}

The deterministic tipping point $(m_R^*, t_c)$ occurs when the positive stable equilibrium is annihilated by the unstable equilibrium: $F(m_R^*, t_c) = 0$ and $\left. \frac{\partial F}{\partial m_R} \right|_{m_R^*} = 0$.

Taking the derivative of Equation \ref{eq:F_m} and applying the phase shift identity $A\sinh(x) + B\cosh(x) = \sqrt{B^2 - A^2}\cosh(x + \phi)$:
\begin{align}
(r_2 - r_1)\sinh(m_R^*) + (r_1 + r_2)\cosh(m_R^*) &= a - J \langle k \rangle \omega \\
2\sqrt{r_1r_2} \cosh(m_R^* + \phi) &= a - J \langle k \rangle \omega
\label{eq:phi}
\end{align}

To derive the phase shift $\phi$, we equate the coefficients from the hyperbolic angle addition formula $\cosh(x + \phi) = \cosh(x)\cosh(\phi) + \sinh(x)\sinh(\phi)$. Matching terms with $A = r_2 - r_1$ and $B = r_1 + r_2$ yields $\tanh(\phi) = A/B$. Thus, the phase shift is inherently an inverse hyperbolic tangent:
\begin{equation*}
\phi = \text{arctanh}\left( \frac{r_2 - r_1}{r_1 + r_2} \right)
\end{equation*}
Using the standard logarithmic definition $\text{arctanh}(z) = \frac{1}{2}\ln\left( \frac{1 + z}{1 - z} \right)$ with $z = (r_2 - r_1)/(r_1 + r_2)$, the numerator becomes $1 + z = 2r_2 / (r_1 + r_2)$ and the denominator becomes $1 - z = 2r_1 / (r_1 + r_2)$. The ratio perfectly simplifies, yielding:
\begin{equation*}
\phi = \frac{1}{2}\ln\left(\frac{r_2}{r_1}\right)
\end{equation*}
Crucially, the real-valued $\text{arctanh}(z)$ is strictly defined only for $z \in (-1, 1)$. Because $r_1$ and $r_2$ represent physical transition rates, they are strictly positive ($r_1 > 0$ and $r_2 > 0$). Therefore, the absolute difference is inherently bounded by the sum, $|r_2 - r_1| < r_1 + r_2$, ensuring that $-1 < z < 1$. This physical constraint guarantees the logarithmic phase shift remains strictly within the real domain.

We define the resilience constant $C$:
\begin{equation} \label{eq:constant_C}
C(\omega) \equiv \frac{a - J \langle k \rangle \omega}{2\sqrt{r_1r_2}}
\end{equation}
Solving for the critical state yields:
\begin{equation} \label{eq:mR_star}
m_R^*(\omega) = \text{arccosh}(C) - \frac{1}{2}\ln\left(\frac{r_2}{r_1}\right)
\end{equation}
For a real saddle node to exist, the necessary condition is $C\geq 1$.
\subsection{Analytical Derivation of the Critical Tipping Point ($t_c$)}
From Eq. \ref{eq:F_m}, and $F(m_R^*, t_c) = 0$, 
\begin{align} 
0 = & (a - J \langle k \rangle \omega)m_R^* - \Big[ (r_2 - r_1)\cosh(m_R^*) + (r_1 + r_2)\sinh(m_R^*) \Big] \nonumber \\
            & - (r_2 - r_1) - k_s\mu + J \langle k \rangle \omega (H_A(0) - vt_c) \nonumber\\
\Rightarrow t_c= & \{(a - J \langle k \rangle \omega)m_R^* - \Big[ (r_2 - r_1)\cosh(m_R^*) + (r_1 + r_2)\sinh(m_R^*) \Big] \nonumber \\
            & - (r_2 - r_1) - k_s\mu + J \langle k \rangle \omega H_A(0) \} /(J\langle k \rangle \omega v)
\end{align}

We know,  
\begin{align}
&\sinh(x + \phi) = \Big[ \sinh(x)\cosh(\phi) + \cosh(x)\sinh(\phi) \Big] \nonumber\\
\Rightarrow &\sqrt{B^2 - A^2} \sinh(x + \phi) = \sqrt{B^2 - A^2} \Big[ \sinh(x)\cosh(\phi) + \cosh(x)\sinh(\phi) \Big] 
\label{eq:tc_1}
\end{align}
From Eq. \ref{eq:phi}, we get $A=r_2-r_1$, $B=r_2+r_1$ and $\tanh(\phi) = A/B$, and using the identity $\cosh^2(\phi) - \sinh^2(\phi) = 1$, we get $\cosh(\phi) = \frac{B}{\sqrt{B^2 - A^2}}$ and $\sinh(\phi) = \frac{A}{\sqrt{B^2 - A^2}}$. Substituting in Eq. \ref{eq:tc_1}, and considering $x=m_R^*$, 
\begin{align}
 &\sqrt{B^2 - A^2} \sinh(m_R^* + \phi) = \Big[ B\sinh(m_R^*) + A\cosh(m_R^*) \Big] \nonumber \\
\Rightarrow & \sqrt{B^2 - A^2} \sinh(m_R^* + \phi)= \Big[ (r_2+r_1)\sinh(m_R^*) + (r_2-r_1)\cosh(m_R^*) \Big] \nonumber \\
\Rightarrow & \Big[ (r_2+r_1)\sinh(m_R^*) + (r_2-r_1)\cosh(m_R^*) \Big]=2\sqrt{r_2r_1}\sinh(m_R^* + \phi) \nonumber \\
\Rightarrow & \Big[ (r_2+r_1)\sinh(m_R^*) + (r_2-r_1)\cosh(m_R^*) \Big]=2\sqrt{r_2r_1}\sinh\Big(\text{arccosh}(C)\Big)\;\;\;\;\;\;\;\;\;\;\;\;[\text from\;\;\ Eq. \ref{eq:mR_star}] \nonumber \\
\Rightarrow & \Big[ (r_2+r_1)\sinh(m_R^*) + (r_2-r_1)\cosh(m_R^*) \Big]=2\sqrt{r_2r_1}\sqrt{C^2 - 1} \;\;\;\;\;\;\;\;\;\;\;\;[\text{As}\;\sinh(\text{arccosh}(C)) = \sqrt{C^2 - 1}]
\label{eq:simple_br}
\end{align}
Using Eqs. \ref{eq:constant_C}, \ref{eq:mR_star} and \ref{eq:simple_br},
 the exact closed-form analytical boundary for deterministic symmetry breaking is at:
\begin{equation} 
t_c(\omega, v) = \frac{2\sqrt{r_1r_2}}{J \langle k \rangle \omega v} \left[ C \text{arccosh}(C) - C \ln\left(\sqrt{\frac{r_2}{r_1}}\right) - \sqrt{C^2 - 1} \right] - \frac{k_s\mu + (r_2 - r_1)}{J \langle k \rangle \omega v} + \frac{H_A(0)}{v}
\label{eq:t_c}
\end{equation}
The intrinsic bias $\mu > 0$ strictly reduces $t_c$, reflecting a lower energy barrier for the noise-driven avalanche.
%
\begin{figure}
         \centering
       \includegraphics[width=\columnwidth]{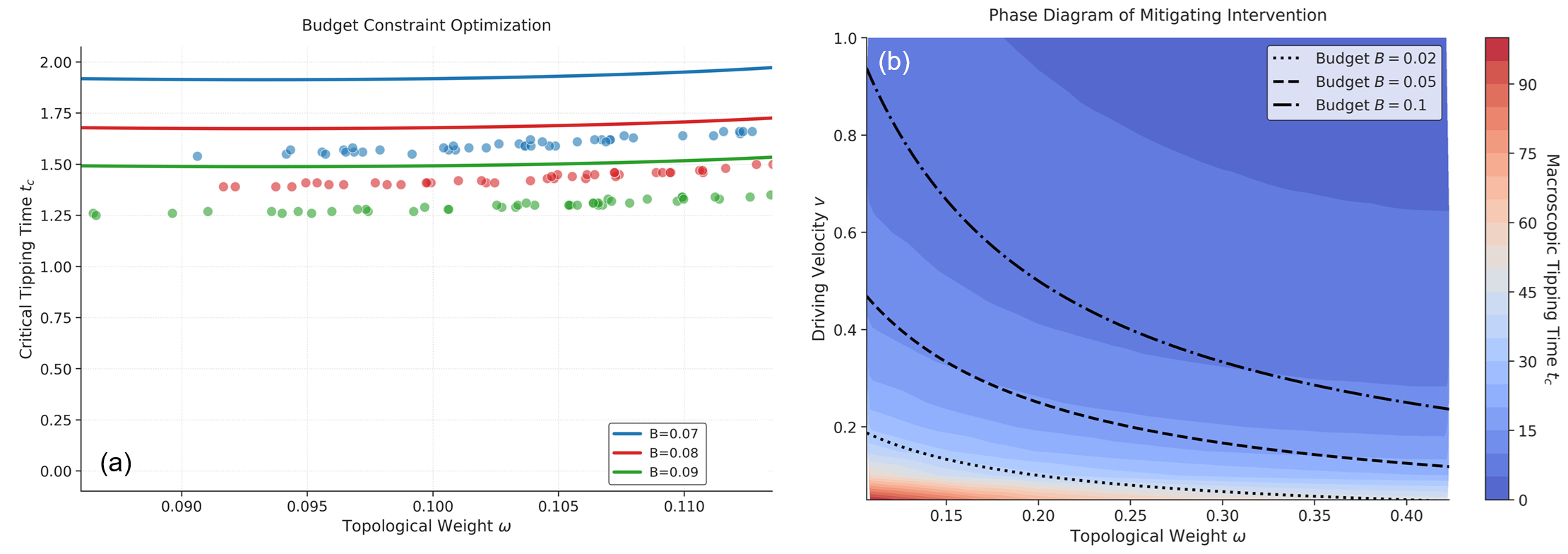}
         \caption{Under intervention budget constraint: (a) Change of tipping time $t_c$ with $\omega$ under fixed budget constraints. The blue, red, and green colors depict theoretical (line) and simulated (circles) values of $t_c$ when $B=0.07, 0.08$, and $0.09$, respectively. (b) Change of $t_c$ (simulated) on $\omega-v$ plane. The iso-cost lines are shown in black. As $\omega$ increases ($v$ decreases), $t_c$ increases over the iso-cost line. Thus, here, a small number of aware nodes with higher $v$ will be more useful to prevent delusional spiraling.  }
         \label{fig:tc_fixedB}
\end{figure}
\subsection{Optimization for The Intervention Budget Constraint}
We impose a fixed \textbf{Intervention Budget} $B$, representing the total topological momentum injected into the network per unit time. We define $B$ as the product of the topological weight and the driving velocity:
\begin{equation}
B = \omega v = \text{const} \implies v = \frac{B}{\omega}
\end{equation}
In other words, we assume the intervention budget is proportional to the product of topological weight and driving speed.

The paradox is thus formulated: To minimize the tipping time $t_c$ under a fixed budget $B$, should one maximize the topological footprint $\omega$ (more nodes, slower velocity) or minimize $\omega$ (few massive hubs, higher velocity)?

We substitute $v = B/\omega$ into Equation (\ref{eq:t_c}) to construct our objective function $t_c(\omega)$. Crucially, $B$ is a fixed, system-wide constant. This substitution isolates $\omega$ as the sole independent variable:
\begin{equation} \label{eq:tc_omega}
t_c(\omega) = \frac{2\sqrt{r_1r_2}}{J \langle k \rangle B} \Phi(C(\omega)) - \frac{k_s\mu + (r_2 - r_1)}{J \langle k \rangle B} + \frac{H_A(0)}{B} \omega
\end{equation}
where 
\begin{equation}
\Phi(C) = C \text{arccosh}(C) - C \ln\left(\sqrt{\frac{r_2}{r_1}}\right) - \sqrt{C^2 - 1}
\label{eq:phi_C}
\end{equation}
To find the optimal strategy, we must calculate the exact gradient $\frac{dt_c}{d\omega}$. Because $B$ is a constant, it behaves as a simple scalar coefficient during differentiation.

Taking the derivative of the objective function (Eq. \ref{eq:tc_omega}) with respect to $\omega$, we get
\begin{align}
\frac{dt_c}{d\omega} =& \frac{2\sqrt{r_1r_2}}{J \langle k \rangle B} \left( \frac{d\Phi}{d\omega} \right) + \frac{H_A(0)}{B} \nonumber \\
=& \frac{2\sqrt{r_1r_2}}{J \langle k \rangle B} \left( \frac{d\Phi}{dC} \frac{dC}{d\omega} \right) + \frac{H_A(0)}{B}
\label{eq:del_tc}
\end{align}
From Eq. \ref{eq:phi_C},
\begin{align}
\frac{d\Phi}{dC}=& \frac{d}{dC} \Big[ C \text{arccosh}(C) \Big]-\frac{d}{dC} \Big[ C \ln\left(\sqrt{\frac{r_2}{r_1}}\right) \Big]- \frac{d}{dC} \Big[ \sqrt{C^2 - 1} \Big] \nonumber\\
= & \text{arccosh}(C) + C \left( \frac{1}{\sqrt{C^2 - 1}} \right)- \frac{1}{2}\ln\left(\frac{r_2}{r_1}\right)- \frac{C}{\sqrt{C^2 - 1}} \nonumber\\
=& \text{arccosh}(C) - \frac{1}{2}\ln\left(\frac{r_2}{r_1}\right) \nonumber\\
=& m^*_R(\omega)
\label{eq:del_phi}
\end{align}
And, from Eq. \ref{eq:constant_C}, we get 

\begin{equation}
\frac{dC}{d\omega} = -\frac{J \langle k \rangle}{2\sqrt{r_1r_2}}
\label{eq:del_C}
\end{equation}

Substituting Eqs. \ref{eq:del_phi} and \ref{eq:del_C} in Eq. \ref{eq:del_tc}, we get
\begin{align}
\frac{dt_c}{d\omega} = & \frac{2\sqrt{r_1r_2}}{J \langle k \rangle B} \Bigg( \Big( m_R^*(\omega) \Big) \left( -\frac{J \langle k \rangle}{2\sqrt{r_1r_2}} \right) \Bigg) + \frac{H_A(0)}{B} \nonumber \\
= & \frac{H_A(0) - m_R^*(\omega)}{B}
\label{eq:grad_tc}
\end{align}

As our overall objective is to minimize $t_c$ such that the society overall aligns with the reality faster, the term $\frac{dt_c}{d\omega}$ largely governs the intervention strategy. The sign of $\frac{dt_c}{d\omega}$ depends upon the difference $(H_A(0)-m_R^*(\omega))$.  If $H_A(0) > m_R^*(\omega)$,  then $\frac{dt_c}{d\omega} > 0$. Because the gradient is strictly positive for all valid $\omega$, the objective function $t_c(\omega)$ is a strictly monotonically increasing function.\\
To minimize the tipping time $t_c$ under a fixed budget $B$, one must minimize $\omega$. Because $v = B/\omega$, this forces $v \to \infty$. This implies that, under this assumption, concentrated, high-velocity interventions outperform distributed, slow interventions in the present model. The algebraic cancellation in the gradient demonstrates that this rule is fundamentally decoupled from the specific network density ($\langle k \rangle$), conformity strength ($J$), or transition rates ($r_1, r_2$). 


\section{Results}
The temporal and spatial evolution of the network's cognitive state is evaluated under both baseline conditions and targeted interventions. Figure 1a illustrates the spatial progression of the node states ($H_i$) on a Barabási-Albert (BA) network. In the baseline scenario lacking aware nodes, the network remains trapped in the delusional basin. However, the introduction of a small targeted fraction of aware hubs ($f=0.05$) forces the macroscopic mean state ($m_R$) to exit the delusional state over time. Figure 1b confirms this macroscopic tipping for the driven mean compared to the baseline. Figure 1c demonstrates that increasing the driving velocity ($v$) of the aware nodes systematically alters the macroscopic tipping trajectory. Figure 1d maps the effect of varying noise intensities ($D$) on the macroscopic mean state, confirming the system's response under different stochastic conditions. \\
The mean-field approximation and the analytically derived critical tipping time ($t_c$) are validated against microscopic Langevin simulations. Figure 2a plots the finite-size scaling of the deterministic tipping time. As the network size ($N$) increases, the gap between the simulated $t_c$ and the theoretical $t_c$ asymptotically decreases for BA networks. This confirms that the theoretical saddle-node bifurcation boundary is rigorous for large social networks in the thermodynamic limit.  Figure 2b displays a parity diagram directly comparing the simulated and analytical $t_c$. Both ER and BA networks tightly follow the diagonal line of perfect agreement. Divergence from the theoretical prediction is observed only when the social conformity parameter ($J$) is drastically reduced, which fundamentally breaks the strong-conformity assumption required for the mean-field reduction. \\
By mapping the diverse network characteristics onto the generalized resilience constant ($C$), the system exhibits universal scaling behavior. Figure 2c demonstrates that the tipping time $t_c$ collapses onto a single master theoretical curve as a function of $C$, regardless of the underlying topology. ER networks, BA networks, and Random Regular networks all closely follow the exact analytical boundary.  The only topology that systematically deviates from this universal curve is the Watts-Strogatz network. This is expected, as its high local clustering coefficient restricts the validity of a global mean-field approximation.\\
Finally, we evaluate the system's non-equilibrium response under a restrictive intervention budget constraint ($B = \omega v$). Figure 3a illustrates the optimization of the critical tipping time $t_c$ with respect to the topological weight $\omega$ across distinct budget tiers ($B = 0.07, 0.08, 0.09$). The solid analytical curves, derived from our saddle-node bifurcation objective function, accurately bound the discrete stochastic simulations on BA networks. As in this simulation $H_A(0)>m_R^*(\omega)$,  following Eq. \ref{eq:grad_tc}, $t_c$ is a strictly monotonically increasing function of $\omega$ along any fixed budget curve. Furthermore, Figure 3b visualizes this optimization landscape as a continuous phase diagram on the $\omega-v$ plane. The heatmap reveals that over an iso-cost line, a network with high topological weight and low driving velocity is trapped in prolonged delusional spiraling, whereas regions of high velocity force rapid macroscopic tipping. By tracing the overlaid dashed iso-cost lines, it becomes visually evident that moving toward a lower $\omega$ and proportionately higher $v$ minimizes the recovery time.

\section{Conclusion}
In this work, we developed a continuous stochastic dynamical framework to model the macroscopic emergence and mitigation of AI-induced delusional spiraling within a socially interacting network. By partitioning the population into a majority of regular nodes driven by conformity and an asymmetric bistable potential, and a minority of unilaterally driving aware hubs, we successfully decoupled the high-dimensional microscopic Langevin equations using a degree-weighted mean-field approximation.\\
Our analytical and computational results yield several core findings. First, we derived a closed-form exact expression for the deterministic critical tipping time ($t_c$) via a saddle-node bifurcation. Finite-size scaling confirms that this theoretical boundary rigorously holds for large networks in the thermodynamic limit. Furthermore, we introduced a fundamental system parameter, the resilience constant $C$, and demonstrated a universal data collapse. Regardless of the underlying topology—whether homogeneous (Erdős-Rényi) or highly heterogeneous (Barabási-Albert)—the macroscopic tipping behavior maps onto a single master theoretical curve. Significant deviations only emerge in topologies with high local clustering, which fundamentally break the mean-field assumption.  \\
Finally, to address mitigation, we evaluated the system under a fixed intervention budget ($B = \omega v$). Rigorous gradient optimization of the objective function $t_c(\omega)$ proves that the system recovery time is strictly monotonically increasing with respect to the topological weight $\omega$. Consequently, concentrated, high-velocity interventions targeting a very small fraction of massive hubs mathematically strictly outperform distributed, slower interventions. This optimal strategy is entirely independent of the overall network density, the strength of social conformity, and the intrinsic potential barrier heights. Overall, this model provides a quantitative, mathematically grounded methodology for understanding how localized, evidence-driven nodes can successfully force a symmetry-breaking transition to rescue a highly connected system from a stable, delusional equilibrium. 

\bibliographystyle{unsrt}
\bibliography{sample}
\end{document}